\documentclass[aps, prl, reprint, showpacs]{revtex4-1}
\usepackage{amsmath}
\usepackage[pdftex]{graphicx}
\usepackage{graphicx}
\usepackage{amsmath}
\usepackage{amssymb}
\usepackage{url}
\usepackage{bm}		
\usepackage{color}					

\begin{document}


\title{Haldane Quantum Hall Effect for Light in a Dynamically Modulated Array of Resonators} 



\author{Momchil Minkov}
\email[]{momchil.minkov@epfl.ch}

\author{Vincenzo Savona}
\affiliation{Laboratory of Theoretical Physics of Nanosystems, Ecole Polytechnique F\'{e}d\'{e}rale de Lausanne (EPFL), CH-1015 Lausanne, Switzerland}

\date{\today}

\begin{abstract}
We study the possibility to induce an effective gauge field for light confined to a Kagom\'{e} lattice of identical optical resonators using an on-site modulation of the resonant frequencies. We find that the Haldane Quantum Hall effect arises simply through a site-dependent phase (but constant amplitude) of the dynamic modulation. Within this scheme, we further demonstrate the existence of topological one-directional edge states immune to back-scattering losses, and discuss the possibilities for a practical implementation, which would enable slow-light devices of unprecedented quality. 
\end{abstract}


\maketitle 

Topological order has opened a new frontier in the classification of distinctive phases of matter, and is thus a center of attention of theoretical and condensed matter physics \cite{Hasan2010}. Its study has also reached the field of photonics \cite{Lu2014}, for two main reasons. First, photonic analogues of topological systems are a promising route to bridging theory and experiment. Second, a signature of a topologically non-trivial material is the presence of one-directional edge states providing energy transport immune to disorder. This could prove extremely valuable for slow-light photonic devices, which find a variety of applications \cite{Krauss2008, Baba2008}, but whose performance is severely limited by back-scattering due to fabrication imperfections \cite{John1987, Patterson2009, Mazoyer2009}.

Historically, topological order was first recognized in relation to the Quantum Hall effect. In that area, Haldane had a ground-breaking contribution in demonstrating that the effect can arise even with zero magnetic field averaged over a primitive cell. The research into topological photonics was also started by Haldane in two theoretical studies \cite{Haldane2008, Raghu2008}, which were quickly followed by an experimental realization of a photonic topological insulator using gyromagnetic media \cite{Wang2009}. This result was however obtained in the GHz frequency range. Due to the lack of suitable materials, reproducing this scheme in the visible or the near-infrared spectrum -- which are the most interesting for applications -- is still a major challenge. The milestone of an experimental realization of topological edge states for light in the near-infrared has been reached using coupled microring resonators \cite{Hafezi2011} or coupled waveguides \cite{Rechtsman2013} by taking advantage of the symmetry-induced degeneracy of rotating and counter-rotating modes. More specifically, these systems are characterized by a preserved time-reversal symmetry (TRS), which leads to an important limitation of the topological protection. The ground-breaking result (which is now known as the Spin Quantum Hall effect) of Kane and Mele \cite{Kane2005} that, for electrons, this protection is still present in TRS systems, relies on the anti-unitarity of the time-reversal operator ($T^2 = -1$). For photons, this operator is unitary, and the result no longer holds \cite{Lu2014}, at least not in its full strength. Instead, the protection relies on the symmetry that prevents the mixing of propagating and counter-propagating modes in a waveguide, which in practice may be broken by disorder. This suggests the need for systems where TRS is broken \cite{Haldane2008, Raghu2008, Koch2010, Umucallar2011, Umucallar2012, Fang2012a, Schmidt2015, Peano2015}. Recently, the  possibility to use a fine-tuned dynamic modulation of a system to engineer a gauge field for photons has been shown both theoretically \cite{Fang2012a} and experimentally \cite{Fang2012, Tzuang2014}. This scheme is employed here to induce a Haldane-like magnetic flux for photons on a lattice of optical resonators.

\begin{figure}
\includegraphics[width = 0.46\textwidth, trim = 0in 0in 0in 0in, clip = true]{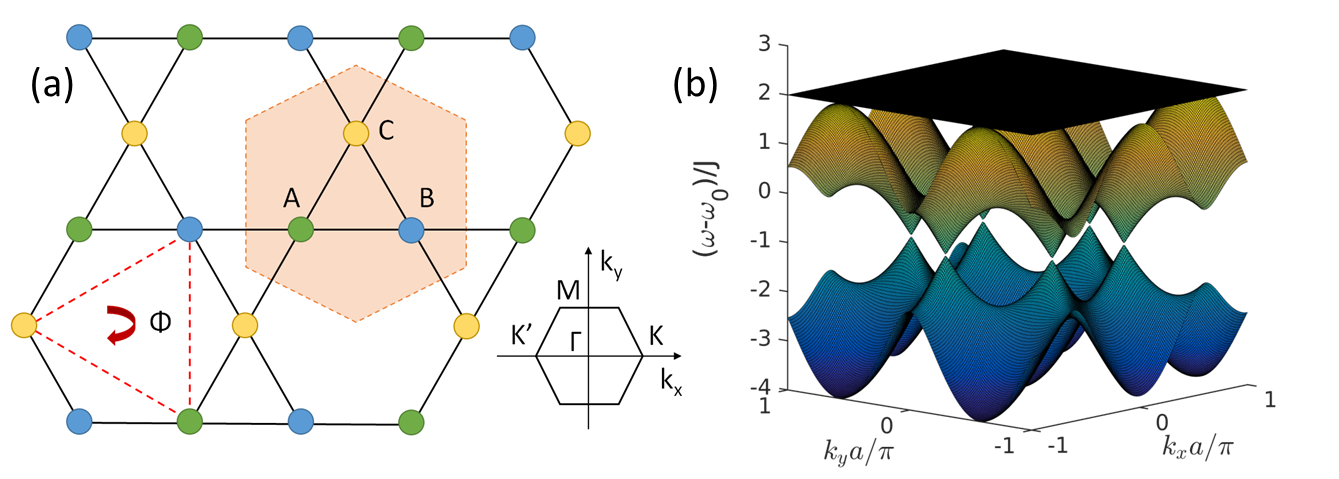}%
 \caption{(a): Kagom\'{e} lattice with three sites in the primitive cell, and the corresponding Brillouin zone. Effective magnetic flux through a hexagon, similar to the one of the original Haldane model, is used here to open topological band gaps. (b): The corresponding band-structure with first-neighbor coupling $J$ and zero flux $\Phi$. There are six Dirac cones, and in addition a flat band.}
\label{fig1}
\end{figure}

\begin{figure*}[ht]
\includegraphics[width = 0.9\textwidth, trim = 0in 0in 0in 0in, clip = true]{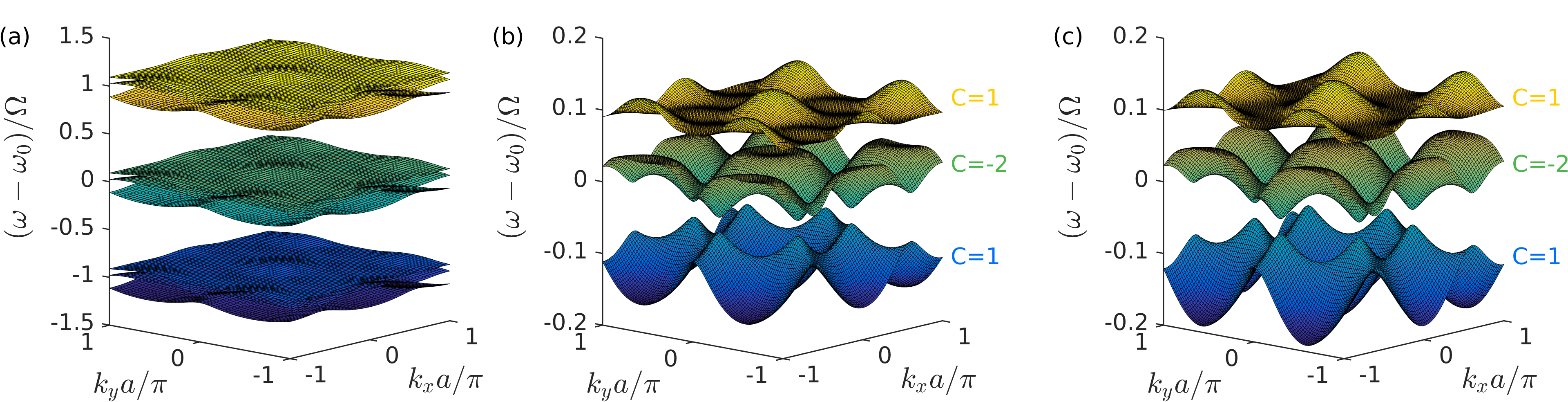}%
 \caption{(a): Quasi-energy bands computed through diagonalization on the Floquet basis, for $J = 0.1\Omega$, $A_0 = 0.9\Omega$, $\varphi = 2.1$. The bands are repeated in orders of $m\Omega$, with $m$ an integer. (b): Zoom-in on the $m = 0$ region of (a). (c): Bands computed through a perturbative expansion of the effective time-independent Hamiltonian. In (b) and (c), the Chern number for each band is indicated.}
\label{fig2}
\end{figure*}

The seminal work by Haldane \cite{Haldane1988} considered a honeycomb lattice (see the Supplemental material) with real first-neighbor and complex second-neighbor couplings. In the absence of the latter, the band structure of the lattice has six Dirac points, and no band gap. Haldane showed that the complex second-neighbor hopping terms, which result in zero average magnetic field over the unit cell, but non-zero magnetic flux through a triangle enclosed by second-neighbor hopping, break the TRS and open a topological band gap. Recently, this was successfully observed in a system of cold atoms in a `shaken' optical lattice \cite{Jotzu2014}, which, together with previous research in that field \cite{Hauke2012, Struck2012, Goldman2014}, inspired the results presented here. In this work, we show how an analogue of the Haldane model can be achieved in a Kagom\'{e} lattice of photonic resonators using a time-periodic modulation of the resonant frequencies, where \textit{only the phase of the modulation varies among different sites, in a spatially periodic manner}. We further show the existence of back-scattering-immune edge states, and discuss the possibilities for a practical implementation of the system.  

\textit{Model--} We consider a lattice of optical resonators, in which the resonant frequencies $\omega_i$ are subject to a periodic modulation in time. The linear photonic Hamiltonian, most generally, reads
\begin{equation}
H = \sum_i (\omega_i + A_i \cos(\Omega t + \phi_i)) a^{\dagger}_i a_i - \sum_{i j} J_{ij} a^{\dagger}_i a_j,
\label{ham_photon}
\end{equation}
where $a^\dagger$ is the photon creation operator, $J_{ij}$ are the hopping coefficients, and $A_i$ and $\phi_i$ denote the site-dependent amplitude and phase of the dynamic modulation, which can be achieved for example through electro-optic modulation \cite{Xu2005, Kuo2005}, optically-induced material non-linearities \cite{Yuce2013}, or optomechanical interaction with phonon modes \cite{Aspelmeyer2014}. The Hamiltonian is particle-number preserving, thus eq. (\ref{ham_photon}) describes the system with any fixed number of photons (sub-spaces of different photon numbers are decoupled). The equation also applies to classical light, since it is a concise way to write the coupled-mode theory that can be used for an array of optical resonators. In the Supplemental material, we outline the theoretical details of the Floquet theory \cite{Shirley1965, Sambe1973, Eckardt2005} that we employ to solve the time-periodic Hamiltonian of eq. (\ref{ham_photon}). The way the Haldane model arises is best revealed through Floquet perturbation theory. There, to first order in $1/\Omega$, new effective couplings can be derived: $J_{ij} \rightarrow J'_{ij} + iJ''_{ij}$, with $J'_{ij}, J''_{ij}$ real constants (see the Supplemental material). In our system, the imaginary $iJ''_{ij}$ thus introduces the magnetic flux required for the Haldane effect.

The most straightforward way to achieve this effect would be by replicating the system of Ref. \cite{Jotzu2014} through an appropriate modulation of a honeycomb lattice of resonators. However, as discussed in the Supplemental material, this would require a spatial gradient of the modulation amplitude $A_i$. This breaks the spatial periodicity and makes it impossible to analyze the system in momentum space, which is a significant theoretical disadvantage. In addition, in view of potential experimental realizations, this feature introduces an extra challenge, since the maximum amplitude of the modulation is inevitably limited, which in turn would limit the maximum system size. Fortunately, this can be easily overcome through a modification of the lattice geometry -- namely, by considering the Kagom\'{e} lattice illustrated in Fig. \ref{fig1}(a). This lattice has three lattice sites per elementary cell, and the band structure (Fig. \ref{fig1}(b)) is similar to the one of the honeycomb lattice in that there are six Dirac cones. The main difference comes from the additional flat band. Importantly, in the presence of a flux similar to the one of the Haldane model, topologically non-trivial band gaps can be opened between the first and the second and/or the second and the third bands \cite{Ohgushi2000, Guo2009}.

The Kagom\'{e} lattice studied here has identical resonators of frequency $\omega_0$ on all sites, and first-neighbor couplings only (along the black lines of Fig. \ref{fig1}(a)), with a hopping coefficient $J$. The dynamic modulation that we assume has the form $\omega_A = \omega_0 + A_0 \cos(\Omega t + \varphi)$, $\omega_B = \omega_0 + A_0 \cos(\Omega t + 2\varphi)$, $\omega_C = \omega_0 + A_0 \cos(\Omega t + 3\varphi)$, where A, B, and C refer to the three sites of the primitive cell. Since the modulation is time-periodic, for time-scales larger than the period $T = 2\pi/\Omega$, we can apply the Floquet theory of quasi-energies \cite{Shirley1965, Sambe1973, Eckardt2005}, which is outlined in detail in the Supplemental material. There, we give two possible ways to compute the Floquet spectrum of the system. This spectrum has a Brillouin-zone-like structure in the sense that, if $\omega$ is an eigenstate of the time-dependent Hamiltonian, so are $\omega + m\Omega$, for all integer $m$. The eigenstates are time-periodic functions with period $T$, and can be expanded on the Floquet basis $| \{n_i\}, m\rangle$, where $n_i$ is the occupation number of site $i$. This results in an infinite-dimensional matrix for diagonalization with matrix elements given in the Supplemental material. We consider the sub-space of a single excitation only, i.e. $n_i = 1, n_{j\neq i} = 0$, and truncate the orders of $m$ at $m_{\mathrm{max}} = 10$ (convergence is always checked), which yields a finite matrix that can be diagonalized numerically. 

The Floquet two-dimensional band structure of the lattice can then be computed in this way, with time- and space-periodic solutions
\begin{equation}
u_n(\mathbf{k}, t) = \sum_{i,m} v_{i, m}(\mathbf{k}, n)e^{-i\mathbf{k}\mathbf{R}_i}e^{i m \Omega t},
\end{equation} 
with $v_{i, m}(\mathbf{k}, n)$ the eigenvectors from the diagonalization, and $\mathbf{R}_i$ the position of site $i$. 
The Floquet band diagram is shown in Fig. \ref{fig2}(a)-(b) for $J = 0.1\Omega$, $A_0 = 0.9\Omega$, $\varphi = 2.1$. As discussed and displayed in panel (a), the bands are repeated in frequency space at an interval of $\Omega$. In panel (b), which shows a close-up of the zero-th order bands of panel (a), we see that band gaps are opened due to the dynamic modulation. To quantify their topological properties, we compute the Chern number for all bands by integrating the Berry curvature $\mathcal{F}(\mathbf{k})$ \cite{Berry1984, Zak1989} over the Brillouin zone. Numerically, we compute $\mathcal{F}(\mathbf{k})$ on a discrete mesh in $\mathbf{k}$-space using the eigenvectors $v_{m, i}(\mathbf{k}, n)$ \cite{Resta2000, Soluyanov2012}. The non-zero Chern numbers (1, -2 and 1 for the three bands, respectively) confirm the non-trivial nature of the band gaps.

The second way to handle eq. (\ref{ham_photon}) (see the Supplemental material) is through a perturbative expansion for an effective time-independent Hamiltonian. As mentioned above, this has the advantage of making the connection between this system and the Haldane model manifest, since the first-order terms in the expansion are imaginary couplings that introduce a flux in the red triangle of Fig. \ref{fig1}(b). In Fig. \ref{fig2}(c), we show the bands computed by diagonalizing this effective Hamiltonian, which agree very well with the exact solution of panel (b), and the computed Chern numbers are the same. 

\begin{figure}
\includegraphics[width = 0.46\textwidth, trim = 0in 0in 0in 0in, clip = true]{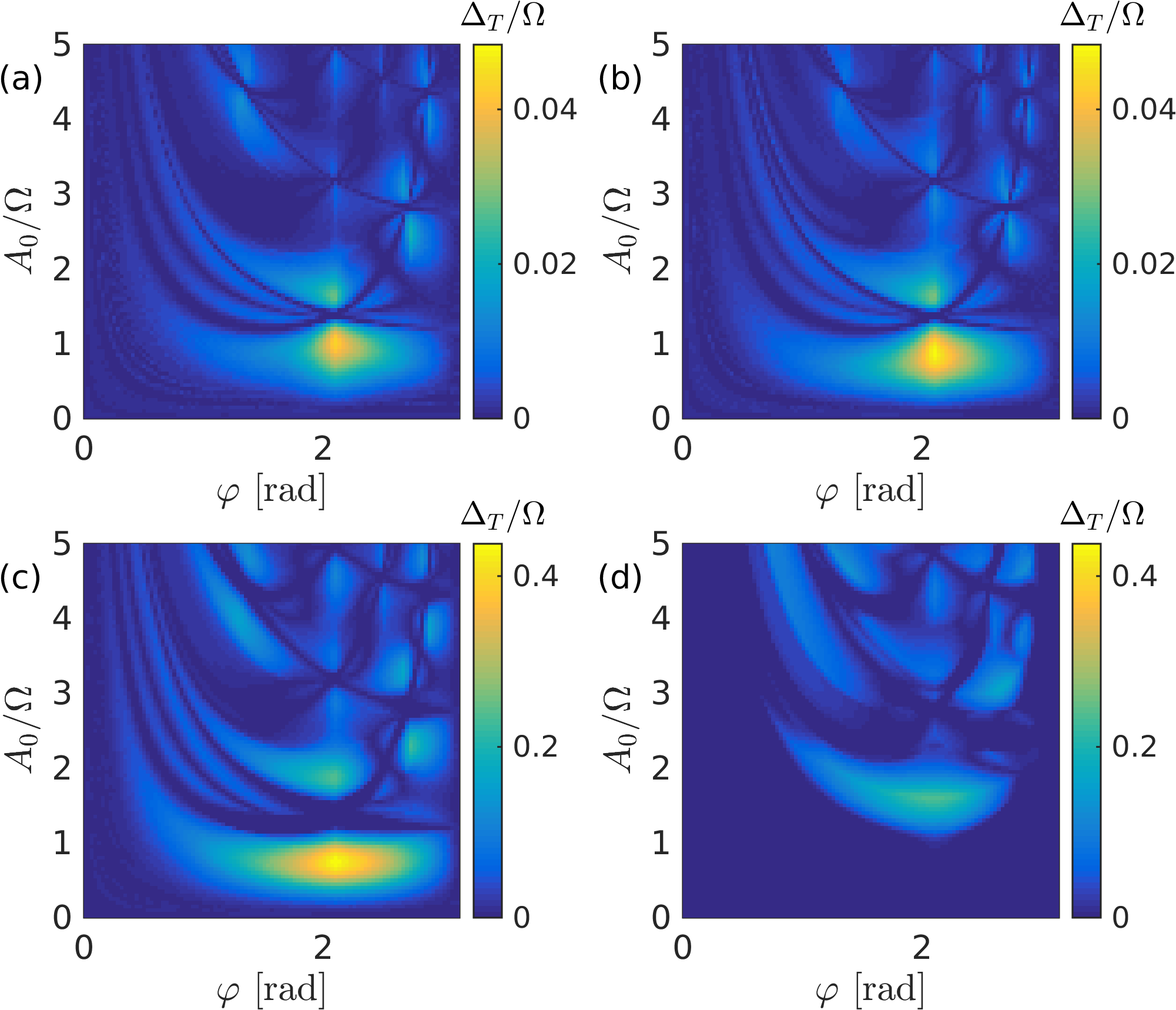}%
 \caption{The (largest) width of the opened band gap due to the dynamic modulation of frequency $\Omega$ vs. the amplitude $A_0$ and the phase angle $\varphi$ for the Kagom\'{e} lattice with first-neighbor coupling $J = 0.1\Omega$, (a): Floquet perturbation theory; (b): expansion on the Floquet basis. (c)-(d): Same as (a)-(b), but for $J = 0.5\Omega$. The color scheme is the same in panels (a) and (b), as well as in panels (c) and (d).}
\label{fig3}
\end{figure}

\begin{figure*}
\includegraphics[width = 0.9\textwidth, trim = 0in 0in 0in 0in, clip = true]{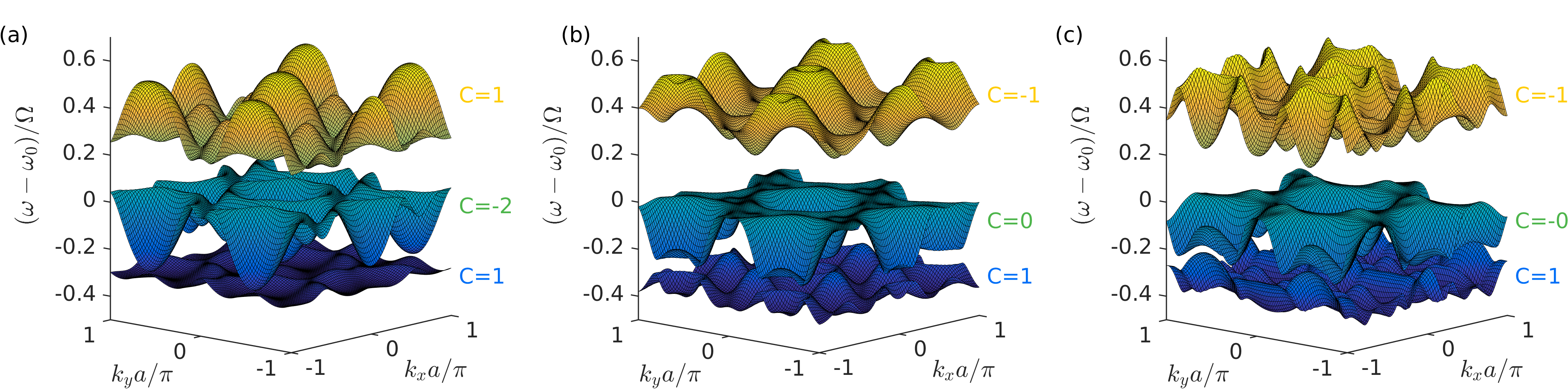}%
 \caption{The bands structure with the largest possible band gap for various values of $J/\Omega$. (a): $J = 0.3\Omega$, $A_0 = 0.5\Omega$, $\varphi = 2.1$; (b): $J = 0.5\Omega$, $A_0 = 1.6\Omega$, $\varphi = 2.1$; (c): $J = 0.7\Omega$, $A_0 = 3.05\Omega$, $\varphi = 2.67$. The Chern number for each band is indicated.}
\label{fig4}
\end{figure*}

\textit{Edge states--} Topological invariants like the Chern number cannot change as long as the band gap remains open. Hence, the width of the band gap is an important parameter, giving an energy scale to the topological protection against disorder (only fluctuations on a larger scale can destroy the topological properties). Thus, in Fig. \ref{fig3}, we plot maps of the gap width $\Delta_T$ (if two gaps are present, the largest value is taken), versus the parameters $A_0$ and $\varphi$. The data in panels (a) and (b) are computed for $J = 0.1\Omega$, with the perturbation theory Hamiltonian in (a), and the full diagonalization in (b), and show very good agreement. In panels (c) and (d), $J = 0.5\Omega$ was used, and the agreement is no longer present. It is natural that the perturbative expansion works well for small $J/\Omega$ when the Floquet bands of different orders are well-separated (Fig \ref{fig2}(a)), but has limited reliability as $J$ increases. Importantly, however, the topological effect is present even beyond perturbation theory: a gap of width larger than $0.2\Omega$ is opened for $J = 0.5\Omega$, $A_0 = 1.6\Omega$, $\varphi = 2.1$. Notice that for any value of the parameters in this system, the band gap is inevitably limited to a fraction of $\Omega$ due to the higher-order Floquet bands. 

In Fig. \ref{fig4}, we show the band structures with the largest band gaps for $J = 0.3\Omega$, $J = 0.5\Omega$ and $J = 0.7\Omega$, with parameters $A_0$ and $\varphi$ chosen for the largest $\Delta_T$ (see Fig. 2 of the Supplemental material). Topologically, there is a difference between the bands in Fig. \ref{fig2}(b) and \ref{fig4}(a), with Chern numbers 1, -2, and 1, and those of Fig. \ref{fig4}(b)-(c) with Chern numbers 1, 0, and -1. What is important, however, is that in both cases there are bands with a non-zero topological invariant. The bulk-boundary correspondence principle \cite{Hasan2010, Lu2014} then applies, guaranteeing the existence of gapless edge states at an interface between the topological material and a topologically trivial one (e.g. empty space). In terms of practical applications, propagating modes robust to disorder are thus expected to appear in a finite system. 

\begin{figure}
\includegraphics[width = 0.42\textwidth, trim = 0in 0in 0in 0in, clip = true]{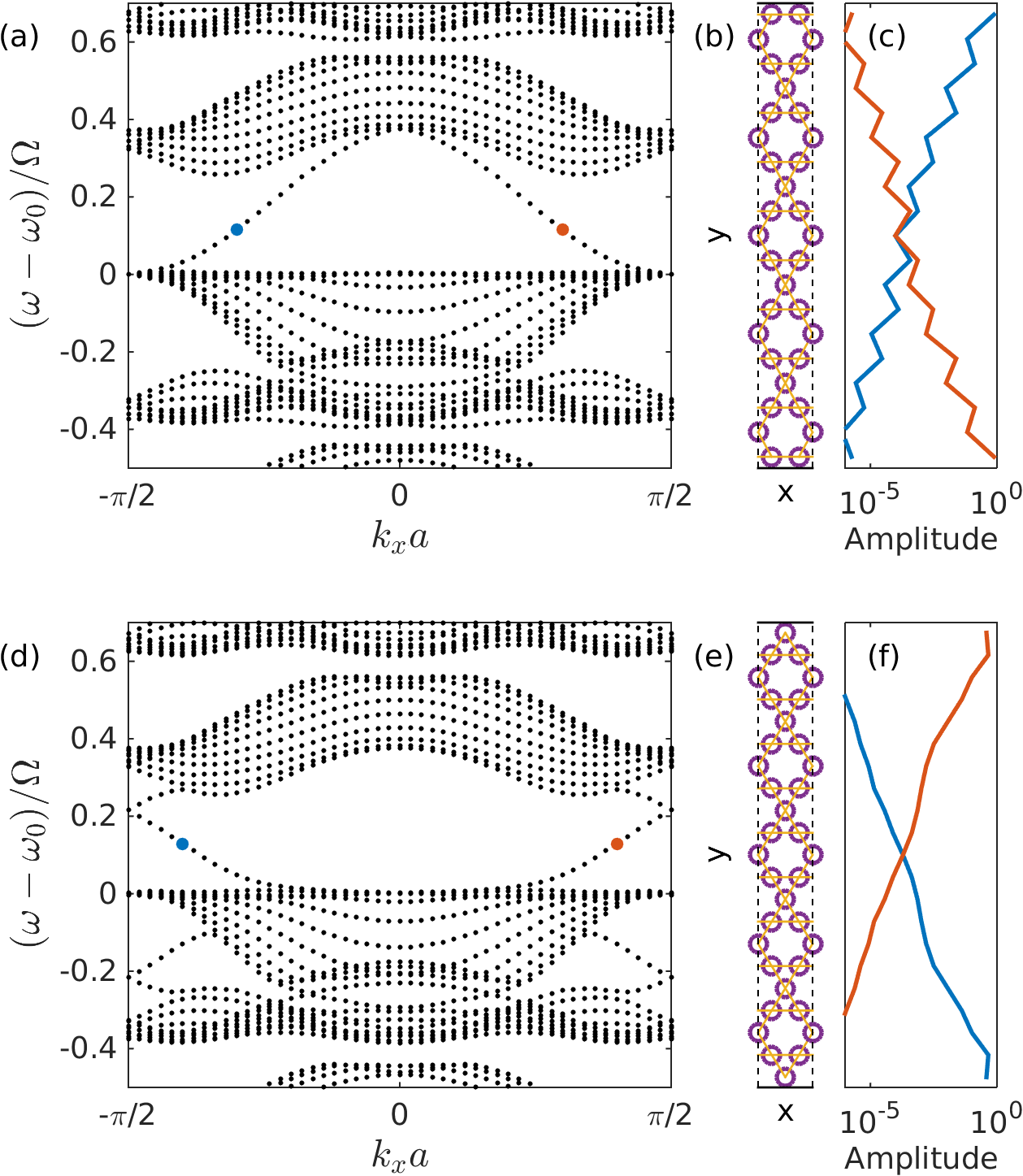}%
 \caption{(a): Floquet bands for the ribbon geometry shown in (b), with a finite number of sites in one direction (the system is truncated at the solid black lines), and periodic boundary conditions in the other (along the dashed black lines). The parameters are as in Fig. \ref{fig3}(b): $J = 0.5\Omega$, $A_0 = 1.6\Omega$, $\phi = 2.1$. (c): The spatial dependence of the eigenstates marked in blue and red, respectively, in panel (a). The y-axis is aligned with the y-axis of panel (b). (d)-(f): Same as (a)-(c), but for a different truncation (compare (b) and (e)).}
\label{fig5}
\end{figure}

The existence of the topological edge modes is illustrated in Fig. \ref{fig5} for a ribbon geometry, with a finite number of sites in one direction, and periodic boundary conditions in the other. The one-dimensional Floquet band structure can again be computed by expanding on the Floquet basis, and is shown in panel (a) and (d) for $J = 0.5\Omega$, $A_0 = 1.6\Omega$, $\varphi = 2.1$. The difference between the two panels comes from the truncation at the edges -- compare panels (b) and (e). Regardless of how we truncate, there is a band that closes the band gap of the bulk structure, due to the non-zero topological invariants. Modes belonging to that band are localized close to the boundaries of the ribbon; the important point, however, is that the modes at $k_x$ and $-k_x$ are localized at opposite edges. This is illustrated in panels (c) and (f), where we plot the position dependence of the magnitude of the eigenvectors of the two states indicated by a blue and a red dot in panels (a) and (d), respectively. The amplitude on the x-axis is the quantity $\sum |v_{m, i}(\mathbf{k}, n)|^2$, where the sum is over all $m$, and over all sites at the same position along y. The edge modes are exponentially localized at the boundaries (notice the logarithmic scale on the x-axes of panels (c) and (f)), thus the overlap between the forward and backward-propagating modes decreases exponentially with the width of the ribbon in the y-direction. This is only possible due to the broken TRS, and ensures protection against back-scattering in the presence of disorder.

\textit{Discussion-- }Several considerations have to be made for the results presented here to have practical implications. We have not considered the loss rate $\kappa$ of the optical resonators, which is in practice always non-zero. To be able to meaningfully talk about light transport, this must be smaller than the coupling constant $J$. In addition, $\kappa$ must also be smaller than the band gap $\Delta_T$, so that the latter can be resolved. By extension, this also implies $\kappa \ll \Omega$. In state-of-the-art photonic crystal cavities, $\kappa/\omega_0$ of the order of $10^{-6}$ can now be routinely achieved \cite{Notomi2008, Lai2014, Sekoguchi2014} at telecommunication frequencies $\omega_0/2\pi \approx 200\mathrm{THz}$, thus $\kappa/2\pi = 0.5\mathrm{GHz}$ is a reasonable and conservative assumption. The coupling constant $J$ is the easiest parameter to control by varying the distance between resonators. Thus the more important challenge is to have a sufficiently high $\Delta_T$. In fact, independently of $\kappa$, $\Delta_T$ is a general figure of merit for the magnitude of the topological protection that should be maximized. 

In Ref. \cite{Fang2012a}, electro-optic modulation was suggested as the practical tool for driving the resonant-frequency oscillation. This offers sufficient control over the phase, and has been shown to be scalable \cite{Fang2012a, Lira2012}. The maximum achievable modulation frequency $\Omega/2\pi$ is of the order of several GHz. A band-gap $\Delta_T$ of the order of $1\mathrm{GHz}$ could thus be achieved, which lies just above the limit set by $\kappa$. We note that this challenge holds both for our proposal and for that of Ref. \cite{Fang2012a}. Very recently \cite{Schmidt2015}, it was suggested to use the coupling of the optical resonators to localized phonon modes to induce the frequency modulation. In this scheme, $\Omega$ is fixed by the phonon resonant frequency, which can be as high as $\Omega/2\pi = 10\mathrm{GHz}$ in two-dimensional optomechanical crystals \cite{Safavi-Naeini2014}. This is sufficiently large for our scheme, and the required phase control can be easily implemented through the phase of the lasers driving the mechanical oscillations \cite{Aspelmeyer2014, Schmidt2015}. We note that, when compared to Refs. \cite{Fang2012a} and \cite{Schmidt2015}, our proposal has a significant structural advantage, as it involves identical resonators with no intermediate (link) elements. Another recent optomechanical scheme \cite{Peano2015} investigated the Kagom\'{e} lattice of resonators, focusing on creating and probing topological states for sound (i.e. phonons). Within that proposal, it is also possible to create topological states of light, but the size of the band gap is shown to be proportional to the phonon hopping coefficient. This is typically orders of magnitude smaller than the phonon resonant frequency, and thus also smaller than the best optical loss rate $\kappa$ that could possibly be achieved in state-of-the-art photonic devices.

While both of the modulation schemes mentioned above could be employed for an experimental realization of our system, a third option is also worth mentioning. Using the optically-induced Kerr nonlinearity, repeated switching at a $\mathrm{THz}$ rate has been recently demonstrated in a micropillar cavity \cite{Yuce2013}. The maximum amplitude in such a scheme is limited to only a fraction of $\Omega$, but assuming $\Omega/2\pi = 1\mathrm{THz}$, $A_0 = 0.05 \Omega$ (which can be read out of the sine-like dependence of the cavity resonant frequency measured in Ref. \cite{Yuce2013}), $J = 0.2\Omega$, and $\varphi = 2.1$, we obtain for our Kagom\'{e} lattice a topological band gap of width $0.033\Omega$, i.e. $\Delta_T/2\pi = 33\mathrm{GHz}$. This value is very similar to the magnitude of the disorder-induced fluctuations in the resonant frequencies of nominally identical photonic crystal cavities \cite{Hagino2009, Minkov2013}, and, furthermore, the latter can in principle be reduced by post-processing techniques \cite{Intonti2012, Piggott2014}. The predicted $\Delta_T$ is thus, on one hand, two orders of magnitude larger than the loss rate of state-of-the-art cavities, and on the other high enough to ensure a truly sizable protection against disorder. 

\textit{Conclusion--} In conclusion, we have described a straightforward implementation of the Haldane-like Quantum Hall effect for light in a lattice of optical cavities, with an effective gauge field produced through a time-periodic modulation of the resonant frequencies. The site-dependence of the phase of the modulation breaks time-reversal symmetry and opens topologically non-trivial band gaps, which, in a finite geometry, yields propagating, back-scattering-free edge states. These can find applications for high bit-rate storage \cite{Baba2008}, for enhanced non-linear effects e.g. for frequency conversion or generation of non-classical light for quantum information processing \cite{Corcoran2009, Monat2010, Azzini2012, Takesue2014}, and for enhanced radiative coupling between distant quantum dots for on-chip quantum computation \cite{Hughes2007, Minkov2013a}.

This work was supported by the Swiss National Science Foundation through Project N\textsuperscript{\underline{o}} 200020\_149537. We thank Hugo Flayac for discussions, and Willem Vos for highlighting the relevance of Ref. \cite{Yuce2013}.  


%

\end{document}